\documentstyle[12pt]{article} \textwidth=17cm \textheight=22.5cm

\def\beq{\begin{equation}} \def\eeq{\end{equation}} \def\bea{\begin{eqnarray}} \def\eea{\end{eqnarray}}
\def\bq{\begin{quote}} \def\eq{\end{quote}}

\def\gappeq{\mathrel{\rlap {\raise.5ex\hbox{$>$}} {\lower.5ex\hbox{$\sim$}}}}

\def\lappeq{\mathrel{\rlap{\raise.5ex\hbox{$<$}} {\lower.5ex\hbox{$\sim$}}}}

\begin{document}
\begin{titlepage}
\vspace*{-1cm}
\phantom{hep-ph/***} 

\hfill{CERN-PH-TH/2005-206}

\vskip 0.5cm
\begin{center}
{\Large\bf The Standard Model of Particle Physics}
\end{center}
\vskip 0.2  cm
\vskip 0.5  cm
\begin{center}
{\large Guido Altarelli}~\footnote{e-mail address: guido.altarelli@cern.ch}
\\
\vskip .1cm
CERN, Department of Physics, Theory Division
\\ 
CH-1211 Geneva 23, Switzerland
\\
\vskip .1cm
and
\\
Dipartimento di Fisica `E.~Amaldi', Universit\`a di Roma Tre
\\ 
INFN, Sezione di Roma Tre, I-00146 Rome, Italy
\\

\end{center}
\vskip 0.7cm
\begin{abstract}
\noindent
A concise introduction to the Standard Model of fundamental particle interactions is presented.
\end{abstract}
\end{titlepage}
\setcounter{footnote}{0}
\vskip2truecm


\section{Introduction} The Standard Model (SM) is a consistent, finite and, within the limitations of our present
technical ability, computable theory of fundamental microscopic interactions that successfully explains most of
the known phenomena in elementary particle physics. The SM describes strong, electromagnetic and weak
interactions. All so far observed microscopic phenomena can be attributed to one or the other of these
interactions. For example, the forces that hold together the protons and the neutrons in the atomic nuclei are due
to strong interactions, the binding of electrons to nuclei in atoms or of atoms in molecules is caused by
electromagnetism and the energy production in the sun and the other stars occurs through nuclear reactions
induced by weak interactions. In principle gravitational forces should also be included in the list of
fundamental interactions but their impact on fundamental particle processes at accessible energies is totally
negligible.

The structure of the SM is a generalisation of that of Quantum Electrodynamics (QED) in the sense that it is a
renormalizable field theory based on a local symmetry (i.e. separately valid at each space-time point $x$) that
extends the gauge invariance of electrodynamics to a larger set of conserved currents and charges. There are
eight strong charges, called "color" charges and four electro-weak charges (which in particular include the
electric charge). The commutators of these charges form the
$SU(3)\bigotimes SU(2)\bigotimes U(1)$ algebra. In QED the interaction between two matter particles with
electric charges, for example two electrons, is mediated by the exchange of one (or more) photons
emitted by one electron and reabsorbed by the second. In the SM the matter fields, all of spin 1/2, are the
quarks, the constituents of protons, neutrons and all hadrons, endowed with both color and electro-weak charges,
and the leptons (the electron $e^-$, the muon $\mu ^-$, the tauon $\tau ^-$ plus the three associated neutrinos
$\nu_e$,$\nu_{\mu}$ and $\nu_{\tau}$) with no color but with electro-weak charges. The matter fermions come in
three generations  or families with identical quantum numbers but different masses. The pattern is as follows:
 \beq
\left[\matrix{
u&u&u&\nu_e\cr
d&d&d&e\cr
} 
\right],~~~~~
\left[\matrix{
c&c&c&\nu_{\mu}\cr
s&s&s&\mu\cr
} 
\right],~~~~~\left[\matrix{
t&t&t&\nu_{\tau}\cr
b&b&b&\tau\cr
} 
\right]. 
\label{123f}
\eeq
Each family contains a weakly charged doublet of quarks, in three color replicas, and a colorless weakly charged
doublet with a neutrino and a charged lepton. At present there is no explanation for this triple repetition of
fermion families. The force carriers, of spin 1, are the photon
$\gamma$, the weak interaction gauge bosons
$W^+$, $W^-$ and $Z_0$ and the eight gluons $g$ that mediate the strong interactions. The photon and the gluons
have zero masses as a consequence of the exact
conservation of the corresponding symmetry generators, the electric charge and the eight color charges. The weak
bosons $W^+$, $W^-$ and $Z_0$ have large masses ($m_W\sim 80.4~GeV$, $m_Z = 91.2~GeV$) signalling that the
corresponding symmetries are badly broken. In the SM the spontaneous breaking of the electroweak gauge symmetry
is induced by the Higgs mechanism which predicts the presence in the physical spectrum of one (or more) spin 0
particles, the Higgs boson(s), not yet experimentally observed. A tremendous experimental effort is underway or
planned to reveal the Higgs sector as the last crucial missing link in the SM verification.

\section{Quantum Chromodynamics}

The statement that QCD is a renormalizable gauge theory based on the group $SU(3)$ with color triplet quark
matter fields fixes the QCD lagrangian density to be:
\beq
{\cal L}~=~-\frac{1}{4}\sum_{A=1}^8F^{A\mu\nu}F^A_{\mu\nu}~+~\sum_{j=1}^{n_f}\bar q_j
(iD\llap{$/$}-m_j)q_j\label{LagQCD}\\
\eeq
Here: $q_j$ are the quark fields (of $n_f$ different flavors) with mass $m_j$; $D\llap{$/$}=D_{\mu}\gamma^{\mu}$, where
$\gamma^{\mu}$ are the Dirac matrices and $D_{\mu}$ is the covariant derivative: 
\beq
D_{\mu}=\partial_{\mu}-ie_s\sum_A~t^Ag_{\mu}^A~;\label{der}\\
\eeq and $e_s$ is the gauge coupling; in analogy with QED:
\beq
\alpha_s=\frac{e_s^2}{4\pi};\label{alfa}\\
\eeq
(in natural units $\hbar=c=1$, always used throughout this summary); $g_{\mu}^A$, $A=1,8$, are the gluon fields
and
$t^A$ are the
$SU(3)$ group generators in the triplet representation of quarks (i.e. $t_A$ are 3x3 matrices acting on $q$); the
generators obey the commutation relations  $[t^A,t^B]=iC_{ABC}t^C$ where $C_{ABC}$ are the complete antisymmetric
structure constants of $SU(3)$ (the normalisation of $C_{ABC}$ and of $e_s$ is specified by
$Tr[t^At^B]=1/2\delta^{AB}$);
\beq
F^A_{\mu\nu}~=~\partial_{\mu} g^A_{\nu}-\partial_{\nu} g^A_{\mu}~-~e_sC_{ABC}g^B_{\mu}g^C_{\nu}.\label{F}\\
\eeq
The physical vertices in QCD include the gluon-quark-antiquark
vertex, analogous to the QED photon-fermion-antifermion coupling, but also the 3-gluon and 4-gluon vertices, of
order $e_s$ and
$e_s^2$ respectively, which have no analogue in an abelian theory like QED. In QED the photon is coupled to all
electrically charged particles but itself is neutral. In QCD the gluons are colored hence self-coupled. This is
reflected in the fact that in QED $F_{\mu\nu}$ is linear in the gauge field, so that the term
$F_{\mu\nu}^2$ in the lagrangian is a pure kinetic term, while in QCD $F^A_{\mu\nu}$ is quadratic in the gauge
field so that in
$F^{A2}_{\mu\nu}$ we find cubic and quartic vertices beyond the kinetic term. 

The QCD lagrangian in eq.(\ref{LagQCD}) has a simple structure but a very rich dynamical content, including the observed complex spectroscopy with a large number of hadrons. The most prominent properties of QCD are asymptotic freedom
and confinement. In field theory the effective coupling of a given interaction vertex is modified by the
interaction. As a result, the measured intensity of the force depends on the tranferred (four)momentum squared, 
$Q^2$, among the participants. In QCD the relevant coupling parameter that appears in physical processes is
$\alpha_s$ (see eq.(\ref{alfa})). Asymptotic freedom means that the effective
coupling becomes a function of $Q^2$: $\alpha_s(Q^2)$ decreases for increasing $Q^2$ and vanishes asymptotically. Thus, the QCD
interaction becomes very weak in processes with large $Q^2$, called hard processes or deep inelastic processes
(i.e. with a final state distribution of momenta and a particle content very different than those in the initial
state). One can prove that in 4 space-time dimensions all gauge theories based on a non commuting group of
symmetry are asymptotically free and conversely. The effective
coupling decreases very slowly at large momenta with the inverse logarithm of $Q^2$:
$\alpha_s(Q^2)=1/b\log{Q^2/\Lambda^2}$ where b is a known constant and $\Lambda$ is an energy of order a few
hundred MeV. Since in quantum mechanics large momenta imply short wavelenghts, the result is that at short
distances the potential between two color charges is similar to the Coulomb potential, i.e. proportional to
$\alpha_s(r)/r$, with an effective color charge which is small at short distances. On the contrary the
interaction strenght becomes large at large distances or small transferred momenta, of order $Q\lappeq
\Lambda$. In fact the observed hadrons are tightly bound composite states of quarks, with compensating color
charges so that they are overall neutral in color. The property of confinement is the impossibility of
separating color charges, like individual quarks and gluons. This is because in QCD the interaction
potential between color charges increases at long distances linearly in r. When we try to separate the quark
and the antiquark that form a color neutral meson the interaction energy grows until pairs of quarks and
antiquarks are created from the vacuum and new neutral mesons are coalesced instead of free quarks. For
example, consider the process $e^+e^- \rightarrow q\bar{q}$ at large center of mass energies. The final
state quark and antiquark have large energies, so they separate in opposite directions very fast. But the
color confinement forces create new pairs in between them. What is observed is two back-to-back jets of
colorless hadrons with a number of slow pions that make the exact separation of the two jets impossible. In
some cases a third well separated jet of hadrons is also observed: these events correspond to the radiation
of an energetic gluon from the parent quark-antiquark pair.

\section{Electroweak Interactions}

We split the electroweak Lagrangian into two parts by separating the Higgs boson couplings:
\begin{equation} {\cal L} = {\cal L}_{\rm symm} + {\cal L}_{\rm Higgs}~.
\label{21}
\end{equation}

We start by specifying ${\cal L}_{\rm symm}$, which involves only gauge bosons and fermions (a sum over all
flavors of quark and leptons, generically indicated by $\psi$, is understood):
\begin{eqnarray} {\cal L}_{\rm symm} &=& -\frac{1}{4}~\sum^3_{A=1}~F^A_{\mu\nu}F^{A\mu\nu} -
\frac{1}{4}B_{\mu\nu}B^{\mu\nu} +
\bar\psi_Li\gamma^{\mu}D_{\mu}\psi_L \nonumber \\ &&+  \bar\psi_Ri\gamma^{\mu}D_{\mu}\psi_R~.
\label{22}
\end{eqnarray} This is the Yang--Mills Lagrangian for the gauge group $SU(2)\otimes U(1)$ with fermion matter
fields. Here
\begin{equation} B_{\mu\nu}  =  \partial_{\mu}B_{\nu} - \partial_{\nu}B_{\mu} \quad {\rm and} \quad F^A_{\mu\nu} =
\partial_{\mu}W^A_{\nu} - \partial_{\nu}W^A_{\mu}  - g \epsilon_{ABC}~W^B_{\mu}W^C_{\nu}
\label{23}
\end{equation} are the gauge antisymmetric tensors constructed out of the gauge field $B_{\mu}$ associated with
$U(1)$, and $W^A_{\mu}$ corresponding to the three $SU(2)$ generators; $\epsilon_{ABC}$ are the group structure
constants, see eq.~(\ref{28}), which, for $SU(2)$, coincide with the totally antisymmetric Levi-Civita tensor
(recall the familiar angular momentum commutators).

The fermion fields are described through their left-hand and right-hand components:
\begin{equation}
\psi_{L,R} = [(1 \mp \gamma_5)/2]\psi, \quad
\bar \psi_{L,R} = \bar \psi[(1 \pm \gamma_5)/2]~.
\label{24}
\end{equation} Note that, as given in Eq. (\ref{24}),
$$
\bar\psi_L = 
\psi^{\dag}_L\gamma_0 = \psi^{\dag}[(1-\gamma_5)/2]\gamma_0 =
\bar\psi[\gamma_0(1-\gamma_5)/2]\gamma_0 = \bar \psi[(1 + \gamma_5)/2]~.
$$ The matrices $P_{\pm} = (1 \pm \gamma_5)/2$ are projectors. They satisfy the relations $P_{\pm}P_{\pm} =
P_{\pm}, P_{\pm}P_{\mp} = 0, P_+ + P_- = 1$.

The standard EW theory is a chiral theory, in the sense that $\psi_L$ and $\psi_R$ behave
differently under the gauge group. In particular, all $\psi_R$ are singlets and all $\psi_L$ are doublets in the
minimal Standard Model (SM). Thus, mass terms for fermions (of the form
$\bar\psi_L\psi_R$ + h.c.) are forbidden in the symmetric limit.  Fermion masses will be
introduced, together with
$W^{\pm}$ and $Z$ masses, by the mechanism of symmetry breaking. The covariant derivatives $D_{\mu}\psi_{L,R}$ are
explicitly given by
\begin{equation} D_{\mu}\psi_{L,R} = 
\left[ \partial_{\mu} + ig \sum^3_{A=1}~t^A_{L,R}W^A_{\mu} + ig'\frac{1}{2}Y_{L,R}B_{\mu} \right] \psi_{L,R}~,
\label{27}
\end{equation}  where $t^A_{L,R}$ and $1/2Y_{L,R}$ are the $SU(2)$ and $U(1)$ generators, respectively, in the
reducible representations $\psi_{L,R}$. The commutation relations of the $SU(2)$ generators are given by
\begin{equation} [t^A_L,t^B_L] = i~\epsilon_{ABC}t^C_L \quad {\rm and} \quad [t^A_R,t^B_R] = i
\epsilon_{ABC}t^C_R~.
\label{28}
\end{equation} We use the normalization $Tr[t^At^B]=1/2\delta^{AB}$ in the fundamental representation of
$SU(2)$. The electric charge generator $Q$ (in units of $e$, the positron charge) is given by
\begin{equation} Q = t^3_L + 1/2~Y_L = t^3_R + 1/2~Y_R~.
\label{29}
\end{equation}

All fermion couplings to the gauge bosons can be derived directly from Eqs. (\ref{22}) and (\ref{27}). The
charged-current (CC) couplings are the simplest. From
\begin{eqnarray} g(t^1W^1_{\mu} + t^2W^2_{\mu}) &=& g \left\{ [(t^1 + it^2)/ \sqrt 2] (W^1_{\mu} -
iW^2_{\mu})/\sqrt 2] + {\rm h.c.} \right\}\nonumber \\
 &= &g \left\{[(t^+W^-_{\mu})/\sqrt 2] + {\rm h.c.} \right\}~,
\label{30}
\end{eqnarray} where $t^{\pm}  = t^1 \pm it^2$ and $W^{\pm} = (W^1 \pm iW^2)/\sqrt 2$, we obtain the vertex
\begin{equation} V_{\bar \psi \psi W}  =  g \bar \psi \gamma_{\mu}\left[ (t^+_L/ \sqrt 2)(1 - \gamma_5)/2 +
(t^+_R/
\sqrt 2)(1 + \gamma_5)/2 \right]
 \psi W^-_{\mu} + {\rm h.c.}
\label{31}
\end{equation}

In the neutral-current (NC) sector, the photon $A_{\mu}$ and the mediator
$Z_{\mu}$ of the weak NC are orthogonal and normalized linear combinations of
$B_{\mu}$ and $W^3_{\mu}$:
\begin{eqnarray} A_{\mu} &=& \cos \theta_WB_{\mu} + \sin \theta_WW^3_{\mu}~, \nonumber \\  Z_{\mu} &=& -\sin
\theta_WB_{\mu} + \cos \theta_W~W^3_{\mu}~.
\label{32}
\end{eqnarray} Equations (\ref{32}) define the weak mixing angle $\theta_W$. The photon is characterized by equal
couplings to left and right fermions with a strength equal to the electric charge. Recalling Eq. (\ref{29}) for
the charge matrix $Q$, we immediately obtain
\begin{equation} g~\sin \theta_W = g'\cos \theta_W = e~,
\label{33}
\end{equation} or equivalently,
\begin{equation} {\rm tg}~\theta_W = g'/g
\label{34}
\end{equation} Once $\theta_W$ has been fixed by the photon couplings, it is a simple matter of algebra to derive
the
$Z$ couplings, with the result
\begin{equation}
\Gamma_{\bar \psi \psi Z} = g/(2~\cos \theta_W) \bar \psi \gamma_{\mu}
  [t^3_L(1-\gamma_5) + t^3_R(1+\gamma_5) - 2Q \sin^2\theta_W] \psi Z^{\mu}~,
\label{35}
\end{equation}  where $\Gamma_{\bar \psi \psi Z}$ is a notation for the vertex. In the MSM, $t^3_R = 0$ and
$t^3_L =
\pm 1/2$. 

In order to derive the effective four-fermion interactions that are equivalent, at low energies, to the CC and NC
couplings given in Eqs. (\ref{31}) and (\ref{35}), we anticipate that large masses, as experimentally observed,
are provided for $W^{\pm}$  and $Z$ by ${\cal L}_{\rm Higgs}$.  For left--left CC couplings, when the momentum
transfer squared can be neglected with respect to
$m^2_W$ in the propagator of Born diagrams with single $W$ exchange, from Eq.~(\ref{31}) we can write
 \begin{equation} {\cal L}^{\rm CC}_{\rm eff} \simeq (g^2/8m^2_W) [ \bar \psi \gamma_{\mu}(1 -
\gamma_5)t^+_L\psi][
\bar \psi
\gamma^{\mu}(1 - \gamma_5) t^-_L\psi]~.
\label{36}
\end{equation}  By specializing further in the case of doublet fields such as $\nu_e-e^-$ or $
\nu_{\mu} - \mu^-$, we obtain the tree-level relation of $g$ with the Fermi coupling constant $G_F$ measured from
$\mu$ decay ($G_F=1.16639(2)~10^{-5}~GeV^{-2}$):
\begin{equation}
 G_F/\sqrt 2 = g^2/8m^2_W~.
\label{37}
\end{equation} By recalling that $g~\sin \theta_W = e$, we can also cast this relation in the form
\begin{equation} m_W = \mu_{\rm Born}/ \sin \theta_W~,
\label{38}
\end{equation} with
\begin{equation}
\mu_{\rm Born} = (\pi \alpha / \sqrt 2 G_F)^{1/2} \simeq 37.2802~{\rm GeV}~,
\label{39}
\end{equation} where $\alpha$ is the fine-structure constant of QED $(\alpha \equiv e^2/4\pi = 1/137.036)$. 

In the same way, for neutral currents we obtain in Born approximation from Eq.~(\ref{35}) the effective
four-fermion interaction given by
\begin{equation} {\cal L}^{\rm NC}_{\rm eff} \simeq \sqrt 2~G_F \rho_0\bar \psi \gamma_{\mu}[...]
\psi \bar \psi \gamma^{\mu}[...] \psi~,
\label{40}
\end{equation} where
\begin{equation} [...] \equiv t^3_L(1 - \gamma_5) + t^3_R (1 + \gamma_5) - 2Q \sin^2\theta_W
\label{41}
\end{equation} and
\begin{equation}
\rho_0 = m^2_W/m^2_Z\cos^2 \theta_W~.
\label{42}
\end{equation}

All couplings given in this section are obtained at tree level and are modified in higher orders of perturbation
theory. In particular, the relations between
$m_W$ and $\sin \theta_W$  [Eqs. (\ref{38}) and (\ref{39})] and the observed values of $\rho~(\rho = \rho_0$ at
tree level) in different NC processes, are altered by computable small EW radiative corrections. 

The gauge-boson self-interactions can be derived from the
$F_{\mu\nu}$ term in ${\cal L}_{\rm symm}$, by using Eq. (\ref{32}) and
$W^{\pm} = (W^1 \pm iW^2)/\sqrt 2$. For the three-gauge-boson vertex $W^+W^-V$ with $V=Z,\gamma$, we obtain
\begin{equation}
\Gamma_{W^-W^+V} = ig_{W^-W^+V}[g_{\mu\nu}(q-p)_{\lambda} + g_{\mu\lambda}(p-r)_{\nu} +
g_{\nu\lambda}(r-q)_{\mu}]~,
\label{43}
\end{equation} with
\begin{equation} g_{W^-W^+\gamma} = g~\sin \theta_W = e \quad {\rm and} \quad g_{W^-W^+Z} = g~\cos \theta_W~.
\label{44}
\end{equation} This form of the triple gauge vertex is very special: in general, there could be departures from
the above SM expression, even restricting us to $SU(2)\otimes U(1)$ gauge symmetric and C and P invariant
couplings. In fact some small corrections are already induced by the radiative corrections. The SM form of the
triple gauge vertex has been experimentally confirmed by measuring the cross-section $e^+e^-\rightarrow W^+W^-$ at
LEP.

We now turn to the Higgs sector of the EW Lagrangian.  The Higgs Lagrangian is specified by the gauge principle
and the requirement of renormalizability to be
\begin{equation} {\cal L}_{\rm Higgs} = (D_{\mu}\phi)^{\dag}(D^{\mu}\phi) - V(\phi^{\dag}\phi) -
\bar \psi_L \Gamma \psi_R \phi - \bar \psi_R \Gamma^{\dag} \psi_L \phi^{\dag}~,
\label{45}
\end{equation} where $\phi$ is a column vector including all Higgs scalar fields; it transforms as a reducible
representation of the gauge group. The quantities $\Gamma$ (which include all coupling constants) are matrices
that make the Yukawa couplings invariant under the Lorentz and gauge groups. The potential $V(\phi^{\dag}\phi)$,
symmetric under $SU(2)
\otimes  U(1)$, contains, at most, quartic terms in $\phi$ so that the theory is renormalizable:
\beq V(\phi^{\dag}\phi)=-\frac{1}{2}\mu^2\phi^{\dag}\phi+\frac{1}{4}\lambda(\phi^{\dag}\phi)^2\label{44a}
\eeq

Spontaneous symmetry breaking is induced if the minimum of V which is the
classical analogue of the quantum mechanical vacuum state (both are the states of minimum energy) is obtained
for non-vanishing $\phi$ values. This occurs because we have taken $\mu^2$ and $\lambda$ positive in V
(note the "wrong" sign of the mass term). Precisely, we denote the vacuum expectation value (VEV) of
$\phi$, i.e. the position of the minimum, by $v$:
\begin{equation}
\langle 0 |\phi (x)|0 \rangle = v \not= 0~.
\label{46}
\end{equation}

The fermion mass matrix is obtained from the Yukawa couplings by replacing $\phi (x)$ by $v$:
\begin{equation} M = \bar \psi_L~{\cal M} \psi_R + \bar \psi_R {\cal M}^{\dag}\psi_L~,
\label{47}
\end{equation} with
\begin{equation} {\cal M} = \Gamma \cdot v~.
\label{48}
\end{equation} In the SM, where all left fermions $\psi_L$ are doublets and all right fermions $\psi_R$ are
singlets, only Higgs doublets can contribute to fermion masses. There are enough free couplings in $\Gamma$, so
that one single complex Higgs doublet is indeed sufficient to generate the most general fermion mass matrix. It
is important to observe that by a suitable change of basis we can always make the matrix ${\cal M}$ hermitian, $\gamma_5$-free 
and diagonal. In fact, we can make separate unitary transformations on $\psi_L$ and $\psi_R$
according to
\begin{equation}
\psi'_L = U\psi_L, \quad \psi'_R = V\psi_R
\label{49}
\end{equation} and consequently
\begin{equation} {\cal M} \rightarrow {\cal M}' = U^{\dag}{\cal M}V~.
\label{50}
\end{equation} This transformation does not alter the general structure of the fermion couplings in ${\cal L}_{\rm
symm}$.

 If only one Higgs doublet is present, the change of basis that makes ${\cal M}$ diagonal will at the same time
diagonalize also the fermion--Higgs Yukawa couplings. Thus, in this case, no flavor-changing neutral Higgs
exchanges are present. This is not true, in general, when there are several Higgs doublets. But one Higgs doublet
for each electric charge sector i.e. one doublet coupled only to $u$-type quarks, one doublet to $d$-type quarks,
one doublet to charged leptons would also be all right, because the mass matrices of fermions with different
charges are diagonalized separately. In fact, at the moment, the simplest model with only one Higgs doublet
seems adequate for describing all observed phenomena.

Weak charged currents are the only tree level interactions in the SM that change flavor: by emission of a W an
up-type quark is turned into a  down-type quark, or a $\nu_l$ neutrino is turned into a $l^-$ charged lepton (all
fermions are letf-handed). If we start from an up quark that is a mass eigenstate, emission of a W turns it into
a down-type quark state d' (the weak isospin partner of u) that in general is not a mass eigenstate. In general,
the mass eigenstates and the weak eigenstates do not coincide and a unitary transformation connects the two sets:
\beq
\left(\matrix{d^\prime\cr s^\prime\cr b^\prime}\right)=V\left(\matrix{d\cr s\cr b}\right)\label{km1}
\eeq or, in shorthand, D'=VD, where V is the Cabibbo-Kobayashi-Maskawa (CKM) matrix. Thus in terms of mass eigenstates the charged weak current
of quarks is of the form:
\beq J^+_{\mu}\propto\bar U \gamma_{\mu}(1-\gamma_5) VD 
\label{km2}
\eeq Since V is unitary (i.e. $VV^\dagger=V^\dagger V=1$) and commutes with $T^2$, $T_3$ and Q (because all
d-type quarks have the same isospin and charge) the neutral current couplings are diagonal both in the primed and
unprimed basis (if the Z down-type quark current is abbreviated as $\bar D^\prime \Gamma D^\prime$ then by
changing basis we get $\bar D V^\dagger \Gamma V D$ and V and $\Gamma$ commute because, as seen from
eq.(\ref{41}), $\Gamma$ is made of Dirac matrices and $T_3$ and Q generator matrices). It follows that $\bar
D^\prime \Gamma D^\prime =\bar D \Gamma D$. This is the Glashow-Iliopoulos-Maiani (GIM) mechanism that ensures natural flavor conservation
of the neutral current couplings at the tree level. For three generations of quarks the CKM matrix depend on four
physical parameters: three mixing angles and one phase. This phase is the unique source of CP violation in the SM.

We now consider the gauge-boson masses and their couplings to the Higgs. These effects are induced by the
$(D_{\mu}\phi)^{\dag}(D^{\mu}\phi)$ term in
${\cal L}_{\rm Higgs}$ [Eq. (\ref{45})], where
\begin{equation} D_{\mu}\phi = \left[ \partial_{\mu} + ig \sum^3_{A=1} t^AW^A_{\mu} + ig'(Y/2)B_{\mu} \right]
\phi~.
\label{51}
\end{equation} Here $t^A$ and $1/2Y$ are the $SU(2) \otimes U(1)$ generators in the reducible representation
spanned by
$\phi$. Not only doublets but all non-singlet Higgs representations can contribute to gauge-boson masses. The
condition that the photon remains massless is equivalent to the condition that the vacuum is electrically neutral:
\begin{equation} Q|v\rangle = (t^3 + \frac{1}{2}Y)|v \rangle = 0~.
\label{52}
\end{equation} The charged $W$ mass is given by the quadratic terms in the $W$ field arising from
${\cal L}_{\rm Higgs}$, when $\phi (x)$ is replaced by $v$. We obtain
\begin{equation} m^2_WW^+_{\mu}W^{- \mu} = g^2|(t^+v/ \sqrt 2)|^2 W^+_{\mu}W^{- \mu}~,
\label{53}
\end{equation} whilst for the $Z$ mass we get [recalling Eq. (\ref{32})]
\begin{equation}
\frac{1}{2}m^2_ZZ_{\mu}Z^{\mu} = |[g \cos \theta_Wt^3 - g' \sin
\theta_W(Y/2)]v|^2Z_{\mu}Z^{\mu}~,
\label{54}
\end{equation} where the factor of 1/2 on the left-hand side is the correct normalization for the definition of
the mass of a neutral field. For Higgs doublets
\begin{equation}
\phi = \pmatrix { \phi^+ \cr
\phi^0}, \quad v = \pmatrix{ 0 \cr v}~, 
\label{56}
\end{equation} we obtain
\begin{equation} m^2_W = 1/2g^2v^2, \quad m^2_Z = 1/2g^2v^2/\cos^2\theta_W~.
\label{58}
\end{equation} Note that by using Eq. (\ref{37}) we obtain
\begin{equation} v = 2^{-3/4}G^{-1/2}_F = 174.1~{\rm GeV}~.
\label{59}
\end{equation} It is also evident that for Higgs doublets
\begin{equation}
\rho_0 = m^2_W/m^2_Z \cos^2\theta_W = 1~.
\label{60}
\end{equation}

This relation is typical of one or more Higgs doublets and would be spoiled by the existence of, for example, Higgs triplets.
These result
is valid at the tree level and is modified by calculable small EW radiative corrections. The $\rho_0$ parameter
has been measured from the intensity of NC interactions (recall eq. (\ref{42})) and confirmed to be close to unity
at a few per mille level.

In the minimal version of the SM only one Higgs doublet is present. Then the fermion--Higgs couplings are in
proportion to the fermion masses. In fact, from the Yukawa couplings $g_{\phi
\bar f f}(\bar f_L \phi f_R + h.c.)$, the mass $m_f$ is obtained by replacing
$\phi$ by $v$, so that $ m_f = g_{\phi \bar f f} v $. In the minimal SM three out of the four Hermitian fields
are removed from the physical spectrum by the Higgs mechanism and become the longitudinal modes of $W^+, W^-$,
and $Z$ which acquire a mass. The fourth neutral Higgs is physical and should be found. If more doublets are
present, two more charged and two more neutral Higgs scalars should be around for each additional doublet.

The couplings of the physical Higgs $H$ to the gauge bosons can be simply obtained from ${\cal L}_{\rm Higgs}$,
by the replacement
\begin{equation}
\phi(x) = \pmatrix{ \phi^+(x) \cr
\phi^0(x)} \rightarrow 
\pmatrix {0 \cr v + (H/\sqrt2)}~,
\label{62}
\end{equation} [so that $(D_{\mu}\phi)^{\dag}(D^{\mu}\phi) = 1/2(\partial_{\mu}H)^2 + ...]$, with the result
\begin{eqnarray} {\cal L} [H,W,Z] &=& g^2(v/\sqrt 2)W^+_{\mu}W^{-\mu} H + (g^2 /4)W^+_{\mu}W^{-\mu}H^2 \nonumber
\\ && + [(g^2vZ_{\mu}Z^{\mu})/(2 \sqrt 2 \cos^2\theta_W)]H \nonumber \\ &&+ [g^2/(8
\cos^2\theta_W)]Z_{\mu}Z^{\mu}H^2~.
\label{63}
\end{eqnarray}

In the minimal SM the Higgs mass $m^2_H\sim \lambda v^2$ is of order of the weak scale v but cannot be predicted
because the value of $\lambda$ is not fixed. The dominant decay mode of the Higgs is in the $b \bar b$ channel
below the WW threshold, while the
$W^+W^-$ channel is dominant for sufficiently large $m_H$. The width is small below the WW threshold, not
exceeding a few MeV, but increases steeply beyond the threshold, reaching the asymptotic value of $\Gamma\sim 1/2
m^3_H$ at large $m_H$, where all energies and masses are in TeV.

A central role in the
experimental verification of the standard electroweak theory has been played by CERN, the European Laboratory
for Particle Physics, located near Geneva, between France and Switzerland. The indirect effects of the $Z_0$,
i.e. the occurrence of weak processes induced by the neutral current, were first observed in 1974 at CERN by the
Collaboration Gargamelle (the name of the bubble chamber used in the experiment). Later, in 1982, always at
CERN, the $W^ \pm$ and the $Z_0$ were for the first time directly produced and observed in
proton-antiproton collisions by the UA1 and UA2 collaborations and then further studied with the
same technique both at CERN and subsequently at the Tevatron of Fermilab near Chicago. Starting
from 1989 LEP, the large $e^+e^-$ collider was functioning at CERN till 2000. In the LEP circular ring of
about 27Km of circumference electrons and positrons were accelerated in opposite directions to an
equal energy in the range between 45 and 103 GeV. The beams were made to cross and collide in
correspondence of four experimental areas where the ALEPH, DELPHI, L3 and OPAL detectors were
located to study the final states produced in the collisions. In its first phase, called LEP1,
from 1989 to 1995 the LEP operation has been completely dedicated at a precise study of the $Z_0$
properties, mass, lifetime, decay modes in order to accurately test the predictions of the SM. The main lesson of the precision tests of the standard electroweak theory can be summarised as follows. It has
been checked that the couplings of quark and leptons to the weak gauge bosons $W^{\pm}$ and $Z$ are indeed
precisely those prescribed by the gauge symmetry. The accuracy of a few $0.1\%$ for these tests implies that, not
only the tree level, but also the structure of quantum corrections has been verified. 
Then, since the end of 1995, the energy of LEP was increased and the phase of LEP2 was
started. The total energy was 
gradually increased up to 206 GeV. The main physics goals of LEP2 were the search for the Higgs and for
possible new particles, the precise measurement of $m_W$ and the experimental study of the triple gauge
vertices $WW\gamma$ and $WWZ_0$. The Higgs particle of the SM could in principle be produced at LEP2 in the
reaction
$e+e-\rightarrow Z_0H$ which proceeds by $Z_0$ exchange. The non observation of the Higgs particle at LEP2 has
allowed to establish a lower limit on its mass: $m_H\gappeq~114~GeV$. Indirect indications on the Higgs mass were
also obtained from the precision tests of the SM, as the radiative corrections effects depend logarithmically on
$m_H$. The indication is that the Higgs mass cannot be too heavy if the SM is valid: $m_H\lappeq~219~GeV$ at
$95\%$ c.l. In 2001 LEP was dismantled and in its tunnel a new double ring of superconducting magnets is being
installed. The new accelerator, the LHC (Large Hadron Collider), will be a proton-proton collider of
$14 ~TeV$ of total center of mass energy. Two large experiments ATLAS and CMS will continue the Higgs hunting
starting in the year 2007. The sensitivity of LHC experiments to the SM Higgs will go up to masses $m_H$ of
about 1 TeV.
 
\section{Keywords}

Fundamental Particle Interactions, Strong, Electroweak, Standard Model, Quarks, Leptons, Gauge Theories, Higgs
Mechanism

\section{Further Reading}

\noindent
Altarelli,G., (2000), "$The Standard~Electroweak~Theory~and~Beyond$", Proceedings of the Summer School on Phenomenology of Gauge Interactions, Zuoz, Switzerland, hep-ph/0011078

\noindent
Altarelli,G., (2001), "$A~QCD~Primer$", Proceedings of the 2001 European School of High Energy Physics, Beatenberg,  Switzerland, hep-ph/0204179.

\noindent
Close,F., Marten, Sutton,M. C., (2002) $"The~Particle~Odyssey"$, Oxford University Press.

\noindent
$"The~Particle~Century"$ ed. by Fraser, G.,(1998), The Institute of Physics. 

\noindent
Martin,B.R., Shaw,G., (1997), ( $"Particle~Physics"$, John Wiley and Sons Ltd; 2nd Ed. 

\noindent
Perkins, D.H., (2000), "$Introduction~to~High~Energy~Physics$. Reading, Usa, Addison-Wesley 

\noindent
Particle Data Group, (2004), Phys. Letters B. 592.

\end{document}